\documentstyle [epsf] {article}
\begin{document}
\begin{flushright}
\quad
\vspace*{-1.8cm}
\flushright{\bf LAL 99-45}\\
\vspace*{-0.3cm}
\flushright{July 1999}
\end{flushright}
\vskip 5 cm
\centerline {\Large\bf AFTER SITGES}
\vskip 4  cm
\centerline {\large\bf F. RICHARD}
\vspace{0.5cm}
\centerline {\Large\bf Laboratoire de l'Acc\'el\'erateur Lin\'eaire, }
\centerline {\it IN2P3-CNRS et Universit\'e de Paris-Sud, BP 34, F-91898 Orsay Cedex, France}
\vskip 5 cm
\noindent
\centerline {\it Summary talk given at the International Workshop on Linear Colliders LCWS 99}
\centerline { Sitges (Barcelona), April 28 - May 5, 1999}
\par
\vskip 4 cm
\vfill\eject

{\bf Introduction}
\vskip 0.5 cm
   This talk is not meant as a summary of the many contributions to this workshop but 
I would rather like to pick up a few key issues which will drive our future work. \par
For the physics issues, one may distinguish
them according to the 2 stages in energy which are foreseen for the next linear collider. \par
Below a centre of mass energy of 500 GeV, one would cover the light Higgs scenario, top physics at threshold (widely studied at
previous workshops) and eventually
take data at the Z pole. The later idea, first proposed by the JLC community, seems a logical step given the
high accuracy which would be achievable on electro-weak measurements with a LC operating at high luminosity with 
longitudinal polarisation. These accuracies would match the precisions which can be obtained for
the top and the Higgs mass, at the level of $\sim$ 100 MeV, with however a 
remaining problem due to the present uncertainty on $\alpha(M_Z)$. \par
 Above a centre of mass energy of 500 GeV, one would cover the realm of SUSY, compositeness (non elementary Higgs scenario, e.g. 
the 
technicolor scenario TC) and indirect searches for Z$^\prime$, new dimensions, anomalous couplings through precision measurements
on fermion pairs and W pairs.
Except for the CLIC scheme (not studied during this workshop), none of the 3 schemes presented so far has the potential to
go above 1 TeV. \par  
 A key issue, which to my mind has not received sufficient attention so far, is to compare systematically
the measurement/discovery potential
of a LC to LHC. This comparison has been initiated in the past and I will recall some relevant results. These comparisons should be
updated taking into account recent studies achieved within the 2 communities. \par
Detector issues will be discussed with attention given to the "Large-Small" debate triggered by the NLC community. Physics issues
are guiding our choices while machine backgrounds and constraints constitute our main limitations. \par
\vskip 0.5 cm
{\bf The light Higgs scenario}
\vskip 0.3 cm
 1/ Prospects for a light Higgs \par
\vskip 0.3 cm
 Precision measurements are consistent with a contribution of a light Higgs\cite{moriond}:
$$ Log_{10}m^{GeV}_H= 1.85^{+0.31}_{-0.39}  $$
 This result is however not a proof of existence of the Higgs boson. Alternatively if there is no Higgs, the SM is
 non-renormalisable
and one may reinterpret this contribution as $Log_{10}\Lambda_{NEW}$ where $\Lambda_{NEW}$ is related to the scale of 
new
physics which provides an adequate U.V. cutoff\cite{alt}. One can therefore conclude optimistically that some 
kind of new physics should appear at
a scale well within the LC energy scale. \par
 The standard model interpretation gives, at 95$\%$ C.L:
$$ 95< m_H < 230~GeV $$
while the central value $\sim$ 100 GeV is consistent with the MSSM 
prediction\cite{hollik}. \par
One should underline once more that a light Higgs scenario appears as an inescapable prediction of SUSY
at low energy scale even within 
non MSSM schemes where there could be substantial reduction of the production cross-section and/or detection problems 
(e.g. invisible decays). The final word on this theory
(the "SUSY killer") therefore
requires\cite{gunion1} a LC operating at centre of mass energies up to 500 GeV. \par
 Alternate schemes are not favoured. \par
The strongly interacting electro-weak scheme would fake a large effective Higgs mass. One can
however conceive that some new physics contribution produces an appropriate cancellation. \par
 The Technicolor scheme, TC, is excluded in its minimal version which mimics the QCD behaviour, since it would give\cite{tak} S$>$0.25.
Precision measurements (assuming a large "Higgs" mass which is appropriate within TC) give\linebreak S=-0.27$\pm$0.12. In extended
versions\cite{lane1}, "walking technicolor", S cannot be computed anymore but the theory predicts light technirhos decaying into terchnipions
which could be found at a LC with clean signatures\cite{lane2} ($b\bar{b}$, $b\bar{c}$, $\tau\nu$...). \par    
\vskip 0.3 cm
 2/ Branching ratios, total width and mass measurements  
\vskip 0.3 cm
    If one can assume a vertex detector starting at a radius of $\sim$ 1 cm 
(see section on detector issues),
the Higgs boson decay modes can be measured with excellent purity/efficiency. In table 1 are indicated the
accuracies which could be reached taking an integrated luminosity of 500 fb$^{-1}$, at $\sqrt{s}$=350 GeV, 
which corresponds to the TESLA scheme. Figure 1 gives the expected variation\cite{battaglia} of these accuracies with
the Higgs mass. 
\begin{table}[t]
\centering
\caption{Expected accuracies on Higgs branching ratios }
\vskip 0.5 truecm
\begin{tabular}{|c|c|c|c|c|c|}
\hline
$b\bar{b}$ & $c\bar{c}$& gg & WW* & $\tau^+\tau^-$ & Inv   \\
\hline 2\%& 8\%  & 6\% & 2-10\% & 6\%& ? \\
\hline
\end{tabular}
\end{table}
 The most challenging mode is clearly $c\bar{c}$ since it requires separation from
the dominant $b\bar{b}$ mode. The separation from the gluon-gluon, gg, channel is also needed
since this mode suffers from large theoretical uncertainties\cite{review}. \par
At this level of experimental accuracy, one needs a special effort to reduce these
uncertainties on $c\bar{c}$ (the running charm mass) and gg 
(this mode is of interest within SUSY since it can be influenced by loop contributions
from the stop sector). \par
It is fair to say that for the $\tau^+\tau^-$ channel full use of the
high accuracy VD has not been achieved and therefore further improvements are expected.
The invisible mode has not yet been investigated but, judging from LEP2 results, it should also lead to
excellent results.
\par
 The branching ratio into WW* becomes measurable if m$_h>$100 GeV. Detailed analyses, in progress\cite{borric}, indicate that
this measurement can be very precise (the range given in table 1 corresponds to a mass interval 110-130 GeV for m$_h$).
 This mode can be used to discriminate between SM and MSSM since the ratios $b\bar{b}$/WW* and $\tau^+\tau^-$/WW* depart
significantly from SM if m$_A<$ 1 TeV. In MSSM the deviation comes from the $b\bar{b}$ and $\tau^+\tau^-$ partial
widths with almost no effect on WW*. One can verify 
this assumption using the Higgstrahlung and fusion processes. For instance $\sigma(hZ)$ can
be measured with an accuracy of $\sim 2\%$. This test would already seem sufficient to most theorists but one can 
also check the h-W-W coupling using 
the fusion process to be complete (there is not
yet an estimate on the accuracy which can be reached on this process). \par    
 If we can measure BR(h$\rightarrow$WW*) and control h-V-V, it becomes possible to derive the total width $\Gamma_T^h$ through the
trivial formula:
$$ \Gamma_T^h=\Gamma_{SM}(h\rightarrow WW*)/BR(h\rightarrow WW*) $$
where I have assumed that h-V-V was found standard. If not, $\Gamma_{SM}(h\rightarrow WW*)$ should be replaced by 
$\Gamma_{SM}(h\rightarrow WW*)\sigma(hZ)/\sigma_{SM}(hZ)$. The precision on $\Gamma_T^h$ follows directly from that on 
BR$(h\rightarrow WW*)$. This level of accuracy, in the mass domain corresponding to MSSM, cannot be achieved with a hadron collider.
\par
The Higgs mass can be determined to better than 100 MeV using the leptonic decays alone and one sees no reason 
which would prevent using as well hadronic decays. An accuracy at the level or better than 50 MeV seems therefore feasible. \par 
This brief panorama of the potential of a LC for what concerns a precise determination of the Higgs properties (mass,
cross-section, branching ratios, total width) illustrates convincingly that a LC can provide unique opportunities for this type 
of physics. \par 
The high luminosity option is necessary to reach the sensitivity needed to distinguish between SM and MSSM in
a domain of parameters which is beyond LHC and LC discovery reach (see figure 2). It seems reasonable to assume that TESLA operating  
at full luminosity for 2 years with an effective integrated time of $\sim$ 10$^7$sec/year could collect 500 fb$^{-1}$ 
at $\sqrt{s}$.  
\vskip 0.3 cm
 3/ Challenging channels
 \vskip 0.3 cm
   The measurement\cite{reid} 
of BR(h$\rightarrow 2 \gamma$) is severely limited by radiative backgrounds and would give a poor accuracy, 
$\sim$ 20$\%$, even requiring integrated luminosities well above 500 fb$^{-1}$. 
An alternate method would be to use a $\gamma-\gamma$
collider which would directly give $\Gamma$(h$\rightarrow 2 \gamma$) with an estimated precision\linebreak 
$\sim$ 3$\%$\cite{roeck}. The total width, obtained
combining the 2 measurements, would therefore also be known at the $\sim$ 20$\%$ level. \par
    The process Zh*, with h*$\rightarrow$2h, would give a direct determination of $\lambda$, the coefficient of the quartic term in
the Higgs potential ($V(\Phi)=\mu^2\Phi^2+\lambda\Phi^4$). At $\sqrt{s}$=500 GeV,
one has typically $\sigma$(Zhh)$\sim$0.3 fb. With an integrated luminosity
of 2 ab$^{-1}$, one could measure $\lambda$ at the $\sim$ 15$\%$ level\cite{kilian}. It is fair to say, however, that present analyses are not yet
optimised and therefore there is room for progress (or reduction on the amount of luminosity needed to achieve this precision). 
This measurement gives a fundamental piece of information on the Higgs potential but how precisely do we need such an information
to be sensitive to non-standard effects ? This question has not, to my knowledge, received a satisfactory answer. \par   
    A similar challenge resides in the measurement of the top-Higgs Yukawa coupling\cite{merino} 
which can be measured from the $t\bar{t}$h final state. This 
cross-section is at the fb level and first estimates are giving a $\sim$ 10$\%$ error on the Higgs Yukawa coupling. \par
    Above channels are clearly at the border of what can be achieved and put severe demands on the quality of the detector
(energy resolution on photons, multi-jet analysis). 
\vskip 0.5 cm
{\bf Physics at $\sqrt{s}>$ 500 GeV}
\vskip 0.5 cm 
 A question naturally arises: can we set a scale on $\sqrt{s}$ for discovering (directly or indirectly) physics beyond the SM ?
So far we can only gather negative informations: \par
 - No significant deviation from SM has been observed so far (if one excepts the super-K results) \par
 - The value of S indicates that we cannot hope for new heavy doublets \par
 - M$_{Z^{\prime}}>$ 600 GeV from the TEVATRON (this limit will rise to $\sim$ 1 TeV with the upgrade) and
 similar limits from LEP2. \par
Supersymmetry appears as the most likely scenario, at least judging from the amount of interest given to this topic in our
community. Unfortunately there is no clear prediction for the relevant mass scale: \par
 - GUT or "naturalness" arguments are too vague since they state that M$_{SUSY}<$1-10 TeV \par
 - Requesting that a LC has the same discovery reach as LHC leads to\linebreak $\sqrt{s}>$ 1.5 TeV (see figure 3) baring in mind that for
these large SUSY scales not covered by a LC at 1 TeV, LHC will need large integrated luminosities and will only perform an 
"inclusive discovery" on the basis of missing transverse energy not being able to make very significant measurements \par
 - This criterion is clearly not true for the chargino-neutralino sector for which a LC with $\sqrt{s}=$ 500 GeV already covers
(see figure 4) a
larger energy domain than LHC (at LHC heavy charginos are only produced through cascades from squarks or gauginos which mean that
their observation is more model dependent) \par
  - Fine-tuning arguments\cite{dimo} are clearly requesting light charginos with the usual ambiguity on the definition of F.T.
(10\% means masses below 90 GeV, 1\% means below 300 GeV) \par
  - EW baryogenesis\cite{caren} calls for a light Higgs and a light stop (LEP2 should cover this scenario). \par  
  In summary, no safe estimate can be made on the minimal energy required for discoveries. One should however remember that
our main concern remains EWSB which, as discussed in previous section, implies that new physics should manifest itself at low scale
unless some malicious cancellation is faking a light Higgs behaviour. If there is no light Higgs one can expect that strong 
interactions will
be observable within the gauge boson sector. This type of scenario will obviously require the highest possible energy 
(see discussion in section on precision measurements). \par
\vskip 0.5 cm
{\bf SUSY}
\vskip 0.5 cm 
 We do not have a SM for SUSY 
but various schemes which are more or less consistent with the constraints imposed by the flavour sector
 (FCNC), fine-tuning, electro-weak symmetry breaking, cosmology: \par
  - mSUGRA\cite{sugr} assumes mass unification for gauginos M$_i$=M$_{1/2}$, and scalars, m$_0$, at GUT scale while $\mu$, the Higgsino 
parameter, is usually related to these masses using the EWSB constraint. 
The LSP is a stable neutralino
(R-parity conservation), a natural candidate for dark matter provided that both m$_0$ and M$_{1/2}$ are not too large 
(therefore within the LC reach) as shown in figure 3.
 \par
  - GMSB\cite{giud} has the same unification assumptions for masses but the LSP is a very light gravitino \par
  - Unification of masses at GUT is not mandatory within "string inspired" models and one may well have unexpected scenarios\cite{gunstr} 
(LSP=gluino if\linebreak M$_3<$M$_{1,2}$, mass degeneracy 
between
the lightest neutralino and the lightest chargino (M$_1$=M$_2$ at our scale) resulting in problematic detections) \par
  - Higher dimensions, at the TeV$^{-1}$ scale, provide a SUSY breaking mechanism which could be compatible with very heavy 
SUSY particles\cite{anto} (up to\linebreak 10 TeV !) with no clear dark matter candidate \par
  - etc... \par
  R-parity violation could be present, providing a mechanism for the super-K effect, 
CP violation\cite{dreesrp}, i.e. mass terms with phases, is possible 
and even desirable in view of baryogenesis. \par
  This brief digest shows that a SUSY scenario could be rather complicated and therefore requires the powerful tools
provided by a LC: well defined energy with polarised beams and clean environment (detector and physics backgrounds). In a
mSUGRA scenario $\mu$, M$_1$ and M$_2$ can be measured independently from 
the neutralino/chargino channels with precisions
typically 10 times better\cite{martyn} than at LHC. If m$_0$ is small enough, a LC allows for very elegant spin-parity measurements 
(A$_{FB}$,A$_{LR}$) which
provide detailed informations: production mechanisms, mass universality in the sleptonic sector, tan$\beta$ etc... \par 
 It is worth noting that there will be more observables than mass parameters and therefore one could be sensitive to the presence of
phases. This has been known for some time\cite{phase} but has not yet been explored quantitatively by experimentalists: an urgent job to 
be done ! \par
 As mentioned previously, life may be very hard if there are degenerate scenarios, e.g. chargino or stop degenerate in mass with
the lightest neutralino. As pointed out\cite{drees} a LC allows to cope with these situations and such techniques have already been
used at LEP2\cite{delphi}(see also the discussion for the Tevatron\cite{gunion}). \par
 The GMSB scenario predicts signatures which require specific features from the detector: \par
- A semi-stable neutralino decaying into photon+gravitino which can be identified as a non-pointing photon with an electro-magnetic
calorimeter segmented longitudinally \par
- A semi-stable slepton decaying into lepton+gravitino which can be identified from the dE/dx information if the tracking device is
gaseous (TPC, drift chamber). \par
 These searches are already performed with LEP detectors. \par
 Additional Higgs bosons H, A, H$^{\pm}$ appearing within MSSM and GMSB 
are predicted heavy within mSUGRA and therefore not necessarily observable at a LC. A significant
 gain in mass range could be achieved by running with a photon-photon collider where single production of neutral Higgs bosons
 allows to reach masses up to 0.8$\sqrt{s}$. Up to now this prospect was not credible given the low cross-section for this
 process and the large background expected. Recently, it was realised that if A and H have masses close enough, 
interference\cite{roeck} could substantially increase the yield and also allow to observe an interesting pattern. \par
  In summary a TeV LC has less discovery range than LHC but leaves no blind regions and gives much more precise and less model 
dependent measurements. A detailed comparison with similar analyses for LHC will help in assessing these statements. \par 
 \vskip 0.5 cm
{\bf Precision Measurements}
\vskip 0.5 cm 
  A future LC will be able to collect very large samples of data and therefore the dominant errors will come from systematics.
A lot will therefore depend on our ability to measure the differential luminosity d$\cal{L}$/d$\sqrt{s}$ (to better than 1\% is
hoped for) and determine precisely the longitudinal 
polarisation. The latter aspect, as recognised by the experts, has so far received little
attention. \par
  The classical menu relies on $f\bar{f}$ and WW final states. 
 \vskip 0.3 cm
  1/ $f\bar{f}$
 \vskip 0.3 cm
    These channels allow to search for various phenomena beyond the SM: $Z^{\prime}$, W$^{\prime}$ (using the $\nu\bar{\nu}$
    channel with ISR), 
Kaluza-Klein excitations, R-parity violation (e.g. s-channel exchange of a virtual sneutrino), leptoquarks (t-channel
exchange), preons etc... Generally speaking one can parametrise these effects in terms of couplings and mass scale and set limits
on these\cite{ruckl}. If some significant deviation is observed, one will in general be able to discriminate between these mechanisms
by using the large set of precise observables (A$_{FB}$,A$_{LR}$,$\tau$ and top polarisation) provided by a LC operating with 
polarised beams. \par
    A LC at $\sqrt{s}$ well below the $Z^{\prime}$ resonance has a discovery reach which extends for $Z^{\prime}$
masses up to $\sim$5-10$\sqrt{s}$ depending on models. Observables allow to discriminate between these various models 
and indirectly estimate M$_{Z^{\prime}}$. On the contrary LHC will be able to provide directly M$_{Z^{\prime}}$ but 
has too few observables to discriminate between the various models. Figure 5 gives a summary of the discovery reach of the various
machines\cite{cvetic}. \par
    A scenario with additional dimensions at a nearby scale has recently received much attention\cite{hewett}.
Schematically, one can distinguish 2 schemes. In practice some models are combining features of these 2 schemes.\par
    In the first scheme one assumes that SM interactions will feel a new dimension at a scale of order TeV$^{-1}$. 
Kaluza-Klein towers
of known particles (e.g. photon, Z) will appear at that scale and modify $f\bar{f}$ observables. \par
    In a second scheme, only gravity sees the new dimensions at a size R which depends on the number of extra dimensions. At this
    scale the Planck mass $M_{Pl}$
is replaced by an effective mass $M_{eff}$ taken $\sim$ 1 TeV to avoid the hierarchy problem
(not below to avoid any conflict with observations). Both masses are related through:
$$  M^2_{Pl}=R^{\delta}M^{2+\delta}_{eff} $$
 where $\delta$ stands for the number of extra dimensions and should be 
above 2 to avoid any conflict with cosmology ($\delta$=2 gives R $\sim$ 1 mm, 
 acceptable for gravity at short distance, but in apparent conflict with supernova observables\cite{super}). \par
  In this scheme Kaluza-Klein excitations of the graviton G are seen near the $M_{eff}$ scale. One can observe
  $e^+e^-\rightarrow\gamma G_{KK}$ which would give extra contribution\cite{kkgam} 
over the genuine $\nu\bar{\nu}\gamma$ background. In passing
  one can note that this signal and the SM background have very similar behaviour in angle and in energy. The sensitivity
  for this channel will therefore depend on our ability to modelise this background (generators) and our control of the 
luminosity. This measurement also requires an efficient coverage of the forward region where most of the signal will be
collected and a veto on forward electrons down to the smallest possible angles to remove photons due to Bhabha scattering.
Studies are under progress\cite{besan} but one can estimate that
these improvement can gain us significant sensitivity and allow to reach limits on $M_{eff}$ at the level of 10 TeV for $\delta$=2
and lower levels for a larger number of dimensions.
\par
 Kaluza-Klein excitations of the graviton 
will also contribute in standard processes giving more model dependent mass limits (with however less
 dependence on $\delta$). The spin 2 of the graviton will reflect on angular dependences. Recently\cite{strum} important constraints
 have been set using the electro-weak precision observables.\par
 At LHC the process $gq\rightarrow q G_{KK}$ will dominate the "monojet" contribution. The sensitivity of LHC will depend on 
systematics (modelling of monojets, calorimetry resolution). It will be interesting to compare the reach of each machine for this
type of new physics. \par
 \vskip 0.3 cm
  2/ WW
 \vskip 0.3 cm
 This channel provides an ideal laboratory to search for anomalous couplings in the gauge sector. As
well known, the sensitivity to anomalies increases dramatically with energy and a gain of 2 orders of magnitude is expected with 
respect to LEP2 as shown \cite{barkww} in figure 6.\par
 Deviations should appear in a strong interaction scenario. The largest effect is observed if there is a nearby
 resonance with I=1 which affects the WW channel (note that I=2 resonances can be accessed with the $e^-e^-$ scheme) as shown
\cite{barkst} in figure 7. 
If there is no such resonance, one can in principle measure some effects predicted in the
general framework of low energy theorems (see figure 7) but this has to be worked out in detail for $\sqrt{s}=$ 1 TeV. \par
 The WW$\nu\bar{\nu}$ channel provides additional informations but seems to give a limited sensitivity for $\sqrt{s}<$ 1 TeV. 
 Typically the precision on anomalous couplings improves by an order of magnitude\cite{zerwas} by increasing $\sqrt{s}$ from 0.8 TeV
to 1.6 TeV. This conclusion should be investigated in more detail, in particular the tradeoff between energy and luminosity and 
limitations due to systematics.           
\vskip 0.5 cm
{\bf Detector issues}
\vskip 0.5 cm 
 This workshop has set very ambitious, but realistic goals for the detector surpassing the LEP/SLD ones. Our aim is to measure
 precisely photons, electrons, muons and jets over the largest possible solid angle. Perfect efficiency is required meaning that
 the detector has to provide excellent pattern recognition (3D for tracking, high granularity also in 3D for calorimetry) with 
 the best possible transparency (low X$^0$). Timing information is also relevant to avoid cumulating mini-jets from different
bunches with however widely different constraints between TESLA (300 ns between bunches) and the warm machines (few ns between
bunches).\par
 Two philosophies have been considered by the NLC community. The Small detector S has a 6 T field, implying that the
 electro-magnetic background due to beam-beam interactions is efficiently trapped and that the Si microvertex 
 detector VD can start
 at R=1cm. A Si tracker is used which guarantees robustness against machine backgrounds. Calorimetry starts at R=80
 cm implying that one can afford very high granularity and excellent energy resolution for photons (Si+W highly segmented).
 In this scenario the last focusing Q-pole can remain outside the coil. \par
 Such a scheme has many obvious advantages with some concern however on the possibility of charge/neutral separation which a
 necessary feature for good energy reconstruction in a magnetic field. \par
 The Large detector L has a gaseous tracker with a radius R $\sim$ 2m allowing good separation between charged and neutrals. 
The gas detector has the advantage of providing high redundancy and allows to measure dE/dx typically with 5\% accuracy
which can be useful for various analyses. It is more vulnerable to machine backgrounds (e.g. positive ions cumulating inside the
sensitive volume of the TPC). The price to pay is of course the large increase of the calorimeter volume which may impose reduced
granularity and therefore poor performances on energy flow and on photon measurements (energy and angle). \par
 A compromise seems possible. A micro-vertex starting at a radius of\linebreak $\sim$ 1 cm is feasible even with a field of 3 T 
in the TESLA scheme with appropriate collimation of the beam. One can therefore assume a solenoidal field with very large radius
which would include the gas tracker and the calorimeters. Fine-grained calorimetry is considered as a working
hypothesis\cite{brient} assuming, for the moment $\sim$ 1cmx1cmx1X$^0$ and Si+W. This option may turn out to be totally unrealistic but the 
proponents of this scheme underline its relevance to reach optimal E-flow performances. Detailed studies are starting for what
concerns gas tracker, showing that they can survive in a LC environment with some caution (e.g. using appropriate R.O. schemes
for the TPC to avoid positive ion return in the sensitive volume, studying gas mixtures without hydrogen to reduce the effect of
slow neutrons). For what concerns the SC Q-pole there does not seem to be any concern for quenches up to a solenoidal field of 4 T.
Clearly all these statements have to be well proven by further R$\&$D studies. \par 
\vskip 0.5 cm
{\bf Detector$\leftrightarrow$Physics}
\vskip 0.5 cm 
 Here I will pick up some physics issues which play a critical part in the design optimisation. \par
\vskip 0.3 cm 
 1/ Charm tagging
\vskip 0.3 cm 
Charm tagging clearly requires a thin VD starting at a radius $\sim$ 1cm. An optimal design should provide 5 $\mu$m on the
asymptotic precision for the impact parameter and about 5 $\mu$m-GeV for the coefficient of the 
scattering term. This can be contrasted with "ordinary"
performances which were taken as\linebreak 10 $\mu$m and 30 $\mu$m-GeV 
respectively. Assuming a charm efficiency at the 50\% level, the gain in
rejection\cite{borp} against beauty is about 4 and the against light quarks of the order of 2.5 as shown in figure 8. This would clearly help in the
charm/gluon/beauty separation needed in the light Higgs scenario but not only. Charm tagging will also be an important tool to tag W 
hadronic decays into c$\bar{s}$ which can be relevant for precision measurements and SUSY analyses (for instance we may want
to determine the sign of a W without requesting a semileptonic decay which 
would reduce kinematical constraints). $\tau$ tagging will
obviously benefit from this improved resolution but to my knowledge this gain has not yet been optimally studied. \par          
The choice of technology (CCD, APS, low resistivity Si) will be ultimately determined from a combination of criteria among which
radiation hardness (electro-magnetic and neutron backgrounds) which has to include a reasonable safety margin, need for a very thin
detector ($\sim$ 0.1\% X$^0$), lowest possible occupancy (which requires an optimal R.O. scheme). Can we stand 1
hit/mm$^2$/Beam-crossing ? The answer is hopefully 
yes provided that we use the outside layers of the VD which have a level of background 
3 orders of magnitude lower to start a precise tracking and assume that we can extrapolate these tracks to sufficient precision
that there are no mis-associations even with such a high occupancy. \par
\vskip 0.3 cm 
 2/ Mini-jets
\vskip 0.3 cm 
 A TPC detector will unavoidably mix interactions occuring in a time slice of order 50 $\mu$s. In the warm scheme this means that the
 whole bunch train is integrated and that several mini-jets will be superimposed on the event. Very precise timing from an other
 device is therefore needed if one wants to separate these minijets which occur within $\sim$ 200 ns. In the TESLA scheme there is
 about 300 ns separation between beam crossing with corresponds to a drift distance of $\sim$ 2 cm, meaning that one can presumably
 separate geometrically tracks occuring from different beam-crossing. For neutrals accurate timing should be provided 
by the calorimeter. 
These aspects, quantified in the JLC studies (possible degradation of Higgs mass reconstruction due to minijets) deserve further 
investigations.
\par
\vskip 0.3 cm 
 3/ Calorimetry
\vskip 0.3 cm 
 E-Flow aspects have been underlined at this workshop with emphasis on the need for calorimeters with good granularity 
allowing to separate
charged and neutrals. One may hope, at best, to achieve a mass resolution of order 30\%/$\sqrt{M_{GeV}}$, which would be about
twice better than for the ALEPH detector. One may however object that at LEP2 hadronic states are reconstructed using  
energy-momentum conservation constraints which provide excellent mass resolution, typically at the 2\% level, i.e. better or
comparable to the best calorimetry that we are discussing. Why should we therefore worry so much to improve on LEP2 performances ?
First one can argue that beamstrahlung will degrade the energy-longitudinal momentum constraint, creating tails. Second one can give
examples where unconstrained channels (e.g. fusion mechanism with 2 spectator neutrinos) require good mass reconstruction (W/Z
separation). Therefore I think that this topic should be given more feedback from physics studies to better assess our need for
the "ultimate" E-flow quality which may be at the limit of reasonable cost. \par         
%\vskip 0.3 cm 
\vfill\eject
4/ Particle ID
\vskip 0.3 cm 
  Electron and muon identification is a basic requirement which can easily be achieved for isolated leptons but may be more 
demanding in complex topologies (6-8 jets) or within jets. A well segmented electro-magnetic calorimeter will be a blessing
to identify non isolated electrons. dE/dx information from the gas detector has proven to be a useful tool for medium energy, $\sim$
5 GeV, electrons. \par
  Search for semi-stable heavy charged sleptons (predicted within GMSB) has been satisfactorily performed at LEP2 
using dE/dx information only. \par
  K-$\pi$ separation can be useful to identify charm and beauty. For instance if the VD has found a secondary vertex consistent
with the charm hypothesis,
 we can tell the sign of the D from the kaon sign (if not we need to ask for a semileptonic). If we want to run high
  statistics at a Z factory, the need for particle identification will be crying... \par
   These arguments, and probably others which have not come to my knowledge, plead for some sort of particle ID. "Poor man's" ID
is clearly dE/dx from a gas detector with 5\% accuracy to provide a modest K-$\pi$ separation. At LEP, dE/dx information
in the ALEPH and DELPHI TPC's was obtained using the wire information which gave important overlap problems (2D information). If
we can imagine a full pad R.O. scheme (3D) without losing the gain uniformity and keep the same amount of sampling, one may 
reach a sufficient dE/dx resolution even within jets. \par
  Cerenkov identification gives clearly a superior K-$\pi$ separation but at the expense of large space losses and, more than
  anything else, material in front of the electro-magnetic calorimeter. One may argue in favour of a DIRC-type solution, as in Babar,
  which is sufficiently compact to reduce above problems but this solution only provides K-$\pi$ separation up to $\sim$ 
4 GeV.  
\vskip 0.3 cm 
 5/ Trigger
\vskip 0.3 cm 
  The issue of triggering has been delicate for LEP experiments, specially for what concerns $\gamma^*-\gamma^*$ physics which does
not allow simple selections. The same is true in the case of SUSY "degenerate" scenarios (e.g. stop mass close to LSP mass)
for which there would be a low energy release. The good news at this workshop is that "there is no trigger at a LC" ! Computing
capabilities are such that we can perfectly hope for a complete registration of our events, with some mild zero suppression to keep
the data volume reasonable, and a full asynchronous treatment of our events between bunch trains which would allow well controlled
selections.   
\vskip 0.5 cm
{\bf The Forward Region}
\vskip 0.5 cm 
 I would like to give a major emphasis to this topic since at future LC forward detection will play a major role. This is clear for
WW, ZZ and $e^+e^-$ processes which give very high rates. 
Tracking and calorimetry should therefore go down to the mask angle, that is $\sim$ 100 mrad. Since the
muon detector extends at a larger distance, muon can even be recovered below this angle.\par
One may worry about the precision on momentum which could be achieved a such angles but in a nice study performed with the S
scheme, it has been shown \cite{schum} that a set of vertical 
Si wheels extending to $\pm$ 150 cm along the beam line, allow to preserve momentum resolution down to the mask
angles (i.e. to better than 10\% up to 250 GeV). 
In the L scheme a similar set up seems feasible (to be checked) provided that the inner radius of the gas detector starts
at $\sim$ 30 cm. 
One could then install a set of such wheels and presumably also have a vertical device behind the end plate of the TPC
to give a precise measurement for energetic tracks (recall however that there should be a very small gap between the end-plate
and the electro-magnetic calorimeter to avoid degradation of the electro-magnetic energy measurement). \par
 The virtue of this forward 
coverage is obviously to reach an almost perfect acceptance($>$99\%). This will be welcome for multijet events for which
acceptance effects go like the power n of the solid angle. Systematics due to acceptance corrections will also 
be very low. One may 
imagine to re-measure at the 
Z pole the ratio R (hadronic rate/$\mu^+\mu^-$) 
which provides a precise determination of $\alpha_s$. Experimental errors, dominated by the
uncertainty on acceptance corrections, would be very much reduced as compared to LEP. \par
 Another major virtue of this almost perfect geometry is to reduce backgrounds induced by the WW process. 
If an electron or a muon from a semileptonic decay is lost, the hadronic decay will look as
 a SUSY candidate. One could argue that mass and energy reconstruction will identify an elastic W
but this is only true on average with tails coming from
 beamstrahlung and from the W Breit-Wigner distribution. \par
  The very forward region also deserves special attention. I have already given an example for "neutrino counting" techniques
  showing that our ability to veto on very forward electrons is essential to improve the sensitivity of this method. One can also 
emphasize SUSY searches which are limited in the "degenerate scenarios" when the mass of the particle comes close to the
mass of the LSP. As an example, if one cannot veto electrons below 50 mrad at $\sqrt{s}$=1 TeV, these searches will be limited by  
$\gamma^*-\gamma^*$ backgrounds for mass differences $\sim$ 25 GeV. At present one is contemplating the possibility to equip the
masks with some active devices and to have a luminosity monitor inside the mask which goes down to 25 mrad. These devices 
can serve as vetoes 
against dangerous $\gamma^*-\gamma^*$ backgrounds, baring in mind that most of the beam-beam
induced background is built of soft particles easy to discriminate. The only potential problem comes from off-momentum 
electrons which, if too
frequent, could generate accidental vetoes at an intolerable rate. Present estimates are that this rate will be reasonable if 
vacuum can be controlled as expected and that only a small fraction of these off-momentum electrons would end their life in the
masking system. More "aggressive" goals, with a veto angle down to 5 mrad are also discussed. \par 
\vskip 0.5 cm
{\bf The Tools}
\vskip 0.5 cm
 Due to lack of time I have been unable to review the tools for physics analyses. I would like however
to make a few short comments on that important topic. \par
 The issue of generators deserves special care since for channels with very high cross-sections like WW, 
$e^+e^-$ and $\nu\bar{\nu}\gamma$, we could be limited by systematics.
From LEP2 we have learned that some items were missing in standard generators
and that the time scale needed to reach a satisfactory situation
was longer than anticipated. Let me give you a few examples. In the GMSB searches we had to
simulate the background producing 2 photons at wide angle and with large 
missing mass. An obvious contribution comes from radiative return to
the Z pole, with Z decaying into $\nu\bar{\nu}$. 2 photon radiation, available,  was insufficient since it left us with a recoil
masses peaking at the Z mass, in disagreement with observations. It is fair to say that no one had asked for 3 hard photon emission
during the LEP2 workshop. $\gamma^*-\gamma^*$ backgrounds, with VDM/QCD/QPM components, are
necessary ingredients for our SUSY searches which
have been rather difficult to control since, for good reasons, they do not motivate the specialists. The QCD sector, where one
needs a good modelling of multijets, is still not completely satisfactory probably for the same reasons. No one is to blame
but we need to get better organized for the next round.\par
 For what concerns simulation and analysis tools, we are clearly still in infancy and we will benefit from the achievements (and
 mistakes !) from Babar and LHC. We should think early on how to keep user friendliness such that a large public, including
 our colleagues theorist, can participate to the preparation of our analyses. I think that this is an essential aspect which has been 
neglected up to now in these big battles to define software products. \par

\vskip 0.5 cm 
\vskip 0.5 cm
{\bf Conclusions}
\vskip 0.5 cm 
 The case for a LC has to be defended not only on grounds of the clean environment and more precise
measurements but there should be a thorough comparison with the LHC potential for physics. \par
The light
Higgs scenario is providing the strongest argument and this has to be consolidated by further progress
both theoretical and experimental. Generally speaking, whether there is or not a light elementary Higgs, 
it seems very probable that the mystery of the EWSB mechanism can be clarified by a TeV LC. \par
The comparison between a TeV LC and LHC on SUSY and precision measurements has to be updated with the
right hypotheses on detector, luminosity and maximal energy. \par
The issue of polarisation, i.e. which amount of polarisation, how precisely known and how it is used, deserves
clarification and detailed work. \par
For the detector/machine aspect, one should decide on the following items:\par

- can we achieve a VD accuracy $\sim$5$\mu$m+5$\mu$m-GeV 
with the Large option ? \par
- can we achieve, at reasonable cost, a fine-grained calorimeter in the Large option and what do we gain on physics ? \par
- down to which angle do we measure (or veto) charged particles, $\mu$, e, $\gamma$ ? \par

 Sitges has given the 3 communities interested in future LC wonderful opportunities to exchange ideas and start discussions on
a wide spectrum of topics. Since large meetings like this one cannot be too frequent,
we should become a world-wide community by continuous communication between individuals sharing the same
interest. \par
 As nicely stated by Michael Peskin in the U.S. meeting, we have a task of "evangelism", that is we have to convince our colleagues
 involved in present collider experiments to join the LC world. Clear physics arguments are therefore needed to enlarge the present
 community and to convince our funding authorities. \par
 There will be 3 projects of TeV collider proposed, corresponding to the 3 regional centers.
 We should help in reaching a rational issue to
 this healthy competition by stating clearly that we think that there should only be one world machine and that we all intend to
 work on it.  
 
\vskip 2.0 cm
\noindent
{\bf Acknowledgements}\par
\vskip 0.5cm
It is a pleasure to thank the organizers for this very nice workshop and to acknowledge useful explanations 
received from many speakers (not to be held responsible of mistakes in this summary). \par

\begin{figure}
\epsfysize15cm
\epsfxsize15cm
\epsffile{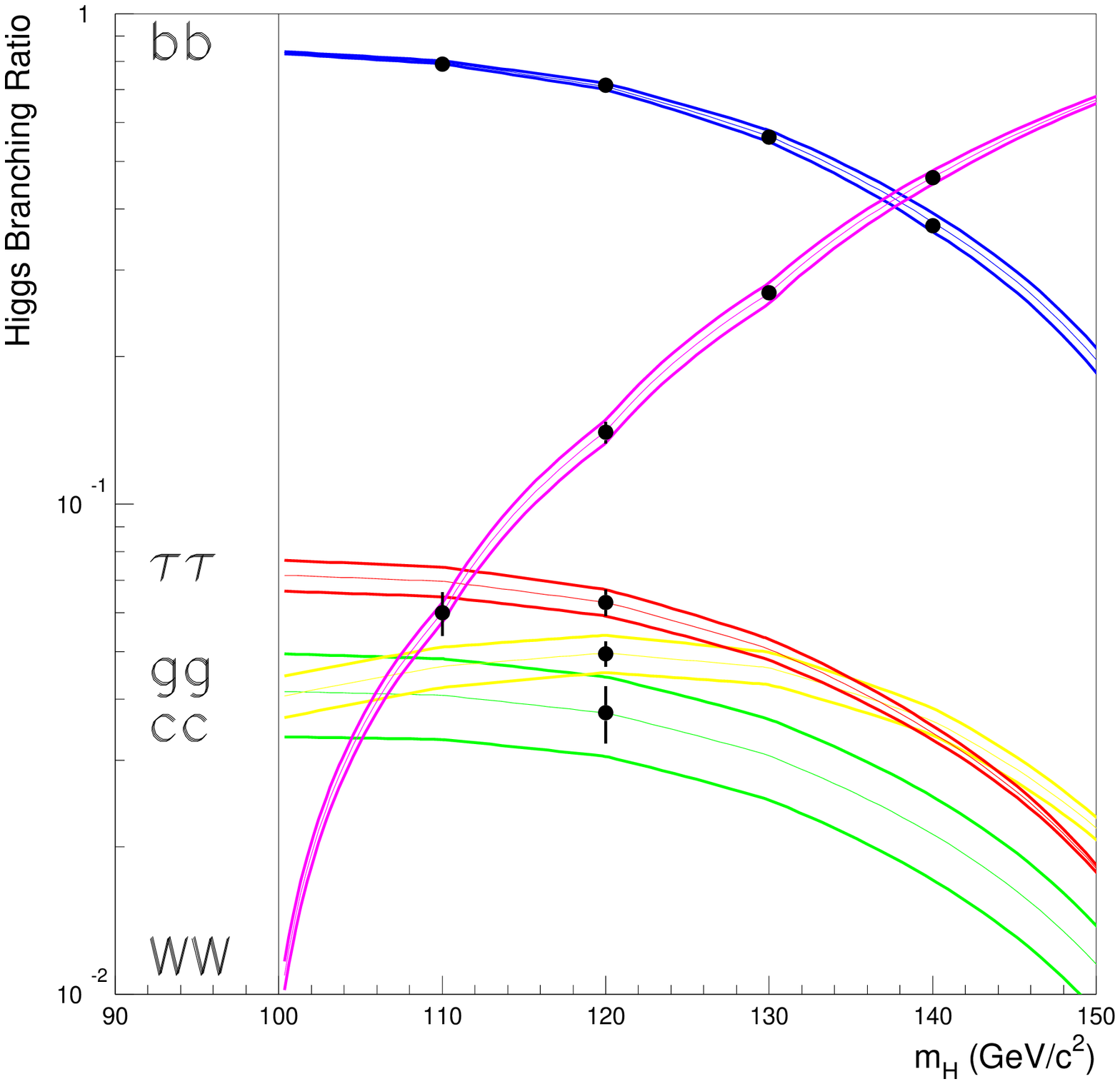}
\caption{Expected precisions on Higgs branching ratios from a LC running at 350 GeV with 500 fb$^{-1}$ versus the Higgs mass. The
dispersion of the curves indicates the theoretical uncertainties.}
\end{figure}
\newpage
\begin{figure}
%\epsfysize16cm
%\epsfxsize15cm
\hspace*{-1.6cm}{\epsffile{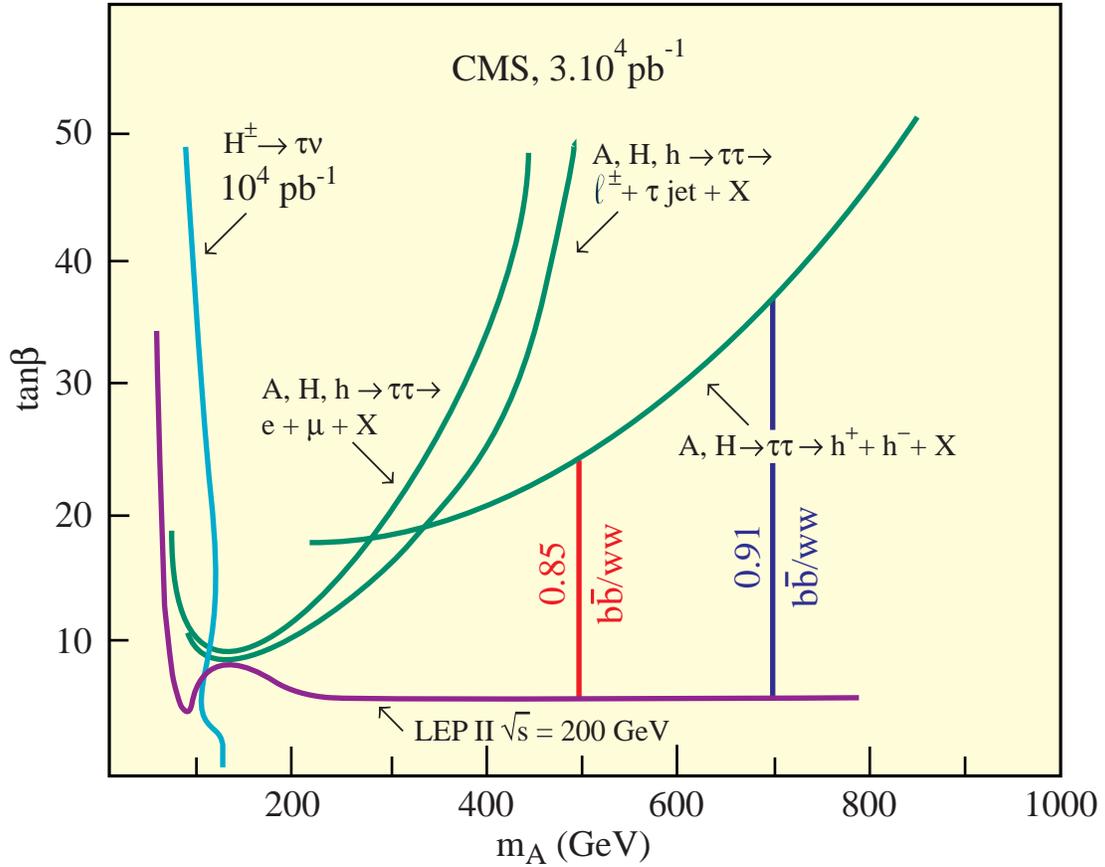}}
\caption{The curves indicate the expected coverage\cite{denegri} of CMS+LEP2 in
the tan$\beta$ m$_A$ plane while the 2 straight lines give the ratio MSSM/SM
of BR(h$\rightarrow$WW*)/BR(h$\rightarrow b\bar{b}$) in a region not excluded by CMS+LEP2.}
\end{figure}
\begin{figure}
\vspace{-1cm}
%\epsfysize17cm
%\epsfxsize14cm
%\hspace*{-1.8cm}{
\epsffile{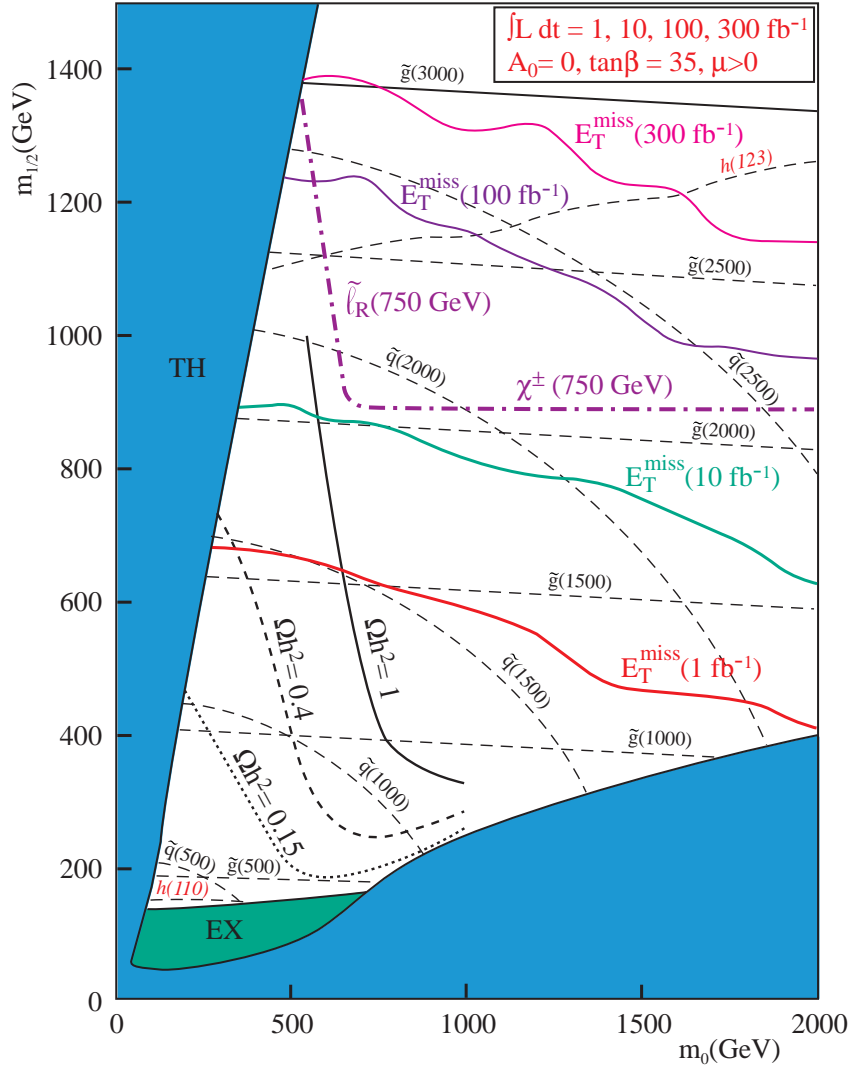}
%\vspace{-0.6cm}
\caption{This figure\cite{denegri} gives the discovery reach of CMS versus integrated luminosity in the MMSM-SUGRA scheme in terms
of the gaugino m$_{1/2}$ and scalar m$_0$ masses adding up all missing energy channels. 
The dashed dotted lines are indicating the reach 
of a LC operating at $\sqrt{s}$=1.5 TeV for right-handed sleptons and for charginos.}
\end{figure}
\newpage
\begin{figure}
\epsfysize11cm
\epsfxsize15cm
\hspace*{-1.5cm}{\epsffile{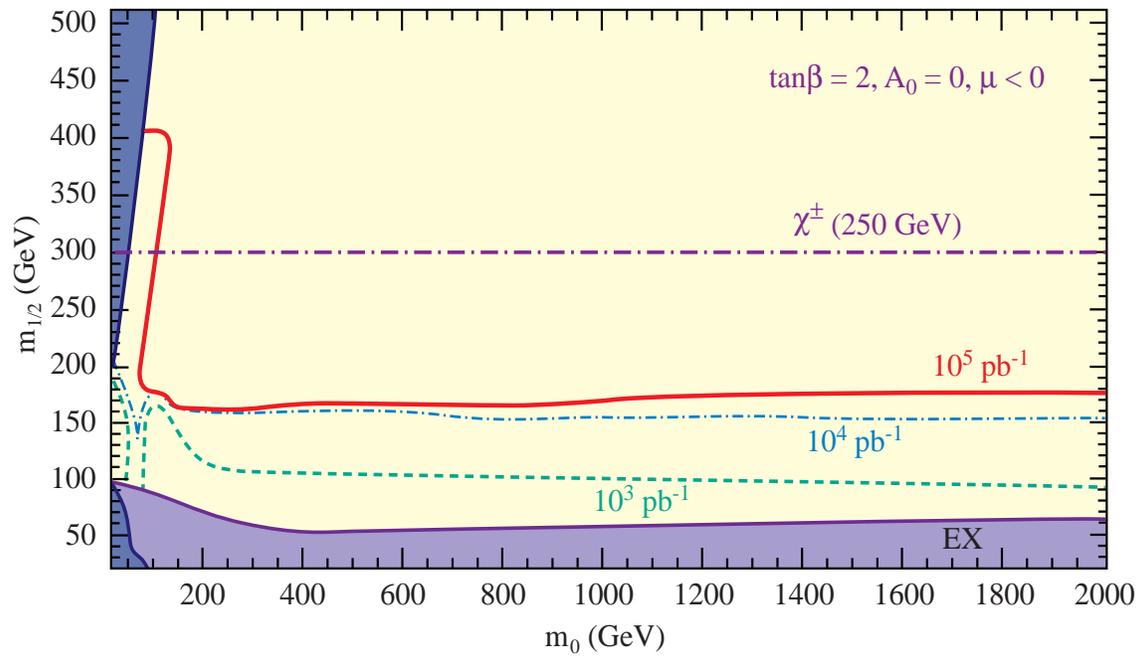}}
\caption{This figure\cite{denegri} gives the discovery reach of CMS versus integrated luminosity in the MMSM-SUGRA scheme in terms
of the gaugino m$_{1/2}$ and scalar m$_0$ masses for the chargino/neutralino searches. The dashed dotted line indicates the reach 
of a LC operating at $\sqrt{s}$=500 GeV for charginos.}
\end{figure}
\begin{figure}
\epsfysize12cm
\epsfxsize12cm
\epsffile{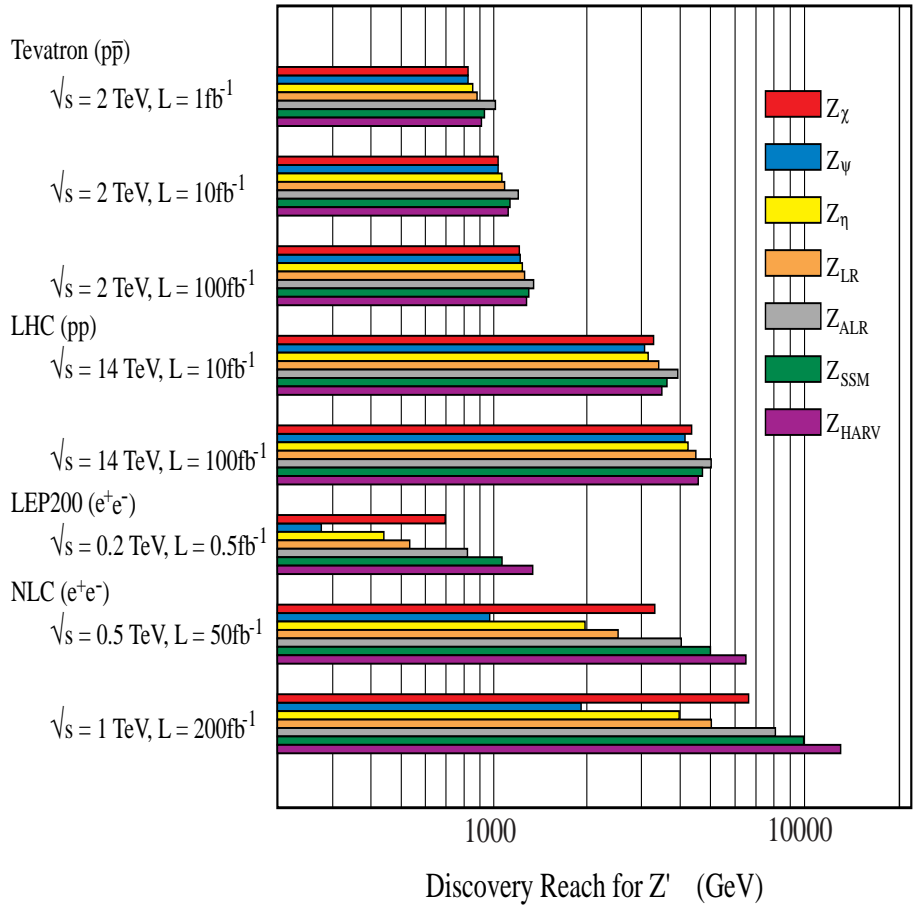}
\caption{Discovery reach for Z$^{\prime}$ given by the various machines for the different scenarios.}
\end{figure}
\begin{figure}
\epsfysize16cm
\epsfxsize16cm
\epsffile{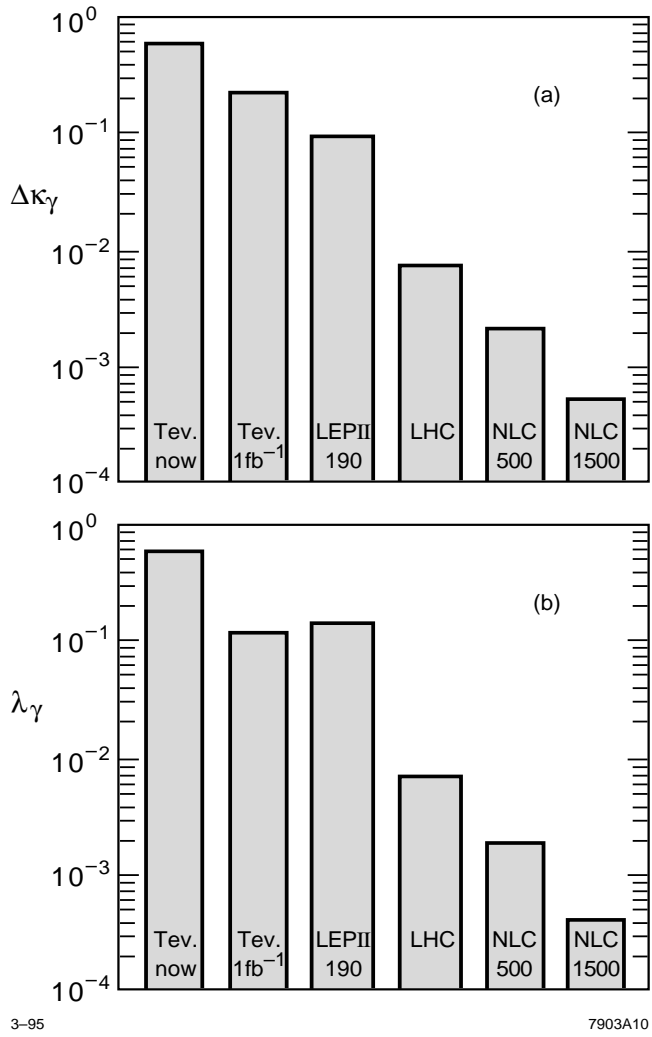}
\caption{Precision on anomalous couplings of photons to W from Tevatron, LEP2 with 0.5 fb$^{-1}$, LHC, 
NLC with 100 fb$^{-1}$ at 500 GeV and with 200 fb$^{-1}$ at 1500 GeV.}
\end{figure}
\begin{figure}
\epsfysize14cm
\epsfxsize14cm
\epsffile{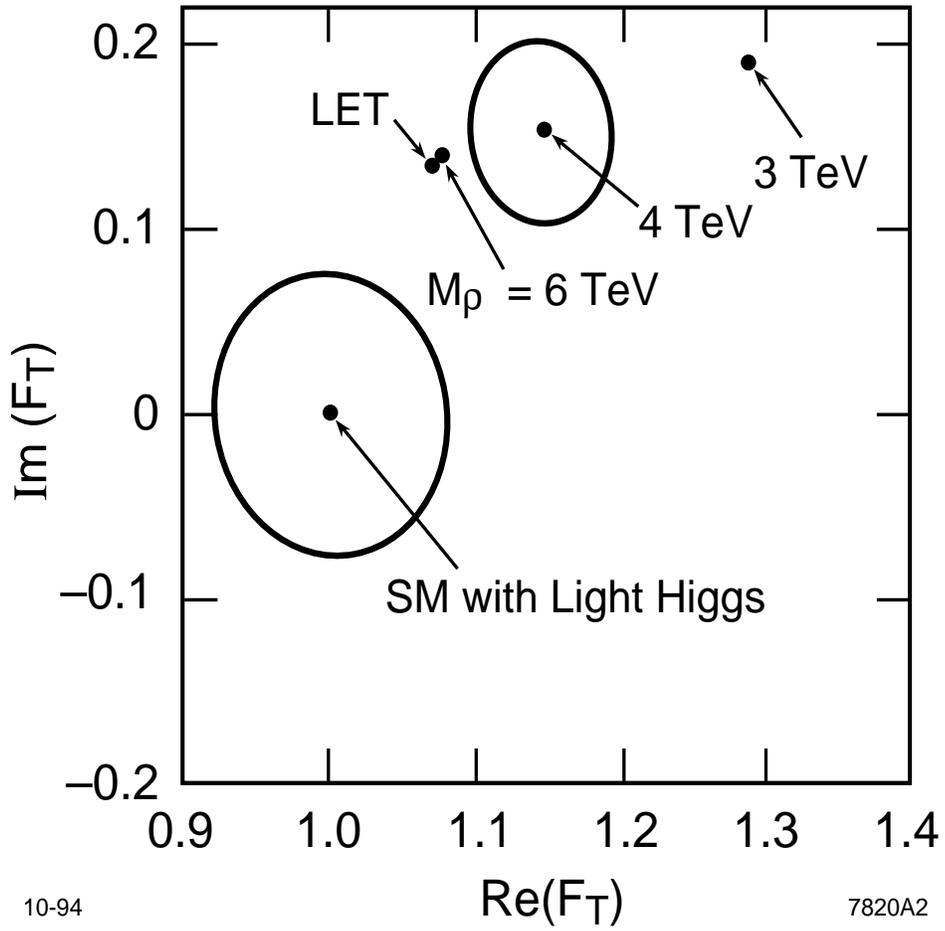}
\caption{Deviations on $e^+e^-\rightarrow W^+W^-$ 
due to strong interactions when there is a $\rho$-type resonance sitting nearby with precisions
(ellipses) given by an NLC at 1.5 TeV and with 200 fb$^{-1}$. 
The point labeled LET corresponds to the deviations without a resonance
(low energy theorems).}
\end{figure}
\begin{figure}
\epsfysize14cm
\epsfxsize14cm
\epsffile{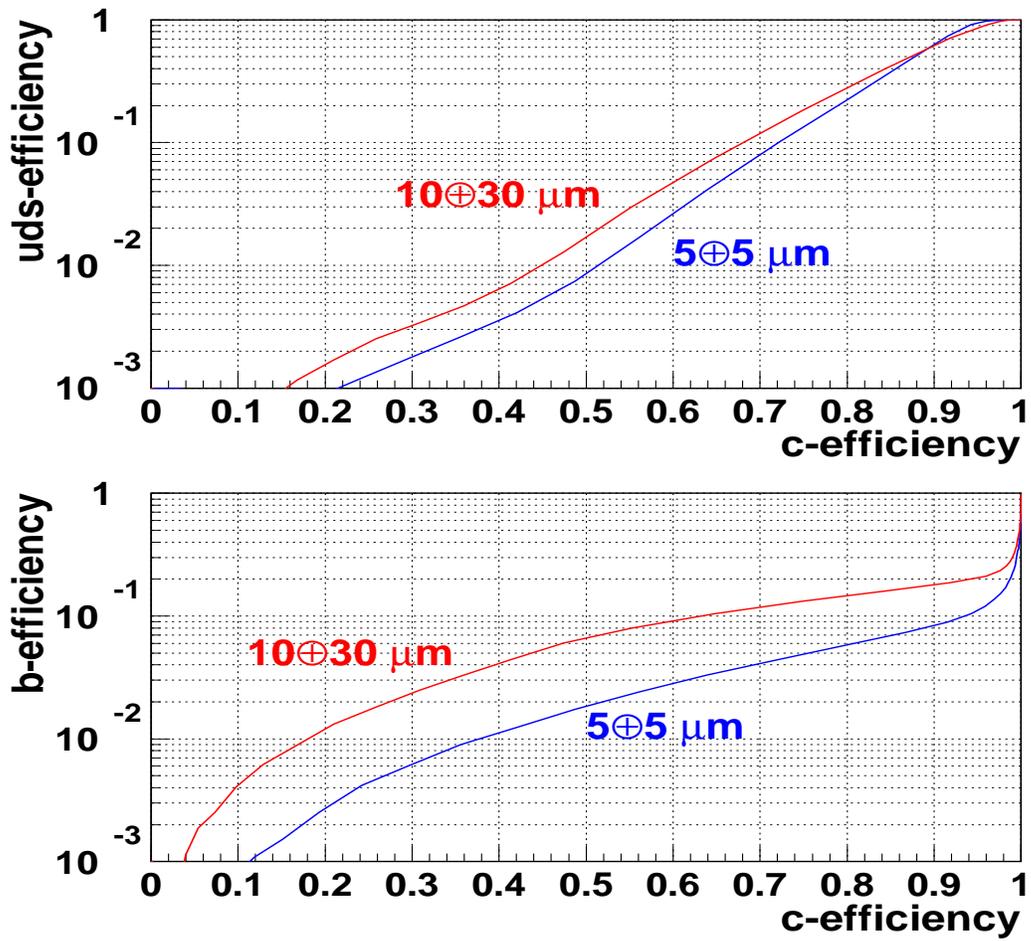}
\caption{The upper plot illustrates the effect of impact parameter
accuracy on light quark(or gluon)/c-quark separability. 
The lower plot gives the information concerning b-quark/c-quark separability.}
\end{figure}
\end{document}